\documentclass[%
 aip,
rsi,%
 amsmath,amssymb,
 reprint,%
]{revtex4-1}

\usepackage{graphicx}
\usepackage{dcolumn}
\usepackage{bm}
\usepackage{notes2bib}
\usepackage{siunitx}

\begin{document}

\preprint{AIP/123-QED}

\title{A low-noise resonant input transimpedance amplified photodetector }

\author{W.~Bowden}
\email[]{william.bowden@npl.co.uk}
\affiliation{National Physical Laboratory, Hampton Road, Teddington TW11 0LW, United Kingdom}

\author{A.~Vianello}
\affiliation{National Physical Laboratory, Hampton Road, Teddington TW11 0LW, United Kingdom}
\affiliation{Imperial College London - Department of Physics, London, SW7 2BB, United Kingdom}

\author{R.~Hobson}
\affiliation{National Physical Laboratory, Hampton Road, Teddington TW11 0LW, United Kingdom}

\date{\today}

\begin{abstract}
We present the design and characterisation of a low-noise, resonant input transimpedance amplified photodetector. The device operates at a resonance frequency of \SI{90}{\mega\hertz} and exhibits an input referred current noise of 1.2 $\textrm{pA}/\sqrt{\textrm{Hz}}$---marginally above the the theoretical limit of 1.0 $\textrm{pA}/\sqrt{\textrm{Hz}}$ set by the room temperature Johnson noise of the detector's \SI{16}{\kilo\ohm} transimpedance. As a result, the photodetector allows for shot-noise limited operation for input powers exceeding \SI{14}{\micro\watt} at \SI{461}{\nano\meter} corresponding to a noise equivalent power of 3.5 $\textrm{pW}/\sqrt{\textrm{Hz}}$. The key design feature which enables this performance is a low-noise, common-source JFET amplifier at the input which helps to reduce the input referred noise contribution of the following amplification stages.

\end{abstract}

\pacs{Valid PACS appear here}
\maketitle

The task of detecting a weak signal encoded in an optical field is ubiquitous across many fields of research. It is often accomplished using a photodiode that produces a photo-current proportional to the signal strength, which is then converted to a voltage with a transimpedance amplifier (TIA). In such applications, it is desirable that the signal-to-noise ratio of the measurement is limited by the inherent shot noise present on the optical signal rather than by electronic noise introduced by the detection electronics. As such, an important figure of merit when choosing a detector is the noise-equivalent-power (NEP), which defines the minimum signal which can be resolved given then detector's intrinsic electronic noise. Along with the NEP, the other important consideration is detector bandwidth. Unfortunately, there is often a trade-off between these two parameters as low noise motivates high transimpedance values, but this limits detector response and, when combined with the parasitic capacitance of the photodiode, leads to increased noise gain at high frequencies \cite{graeme1996}. 

One approach to overcome this limitation is to adopt a resonant architecture. By adding an inductive element at the input, the effect of parasitic capacitance can be suppressed, enabling a large transimpedance gain at a specific resonance frequency. As such, they are well-suited to heterodyne and homodyne measurement schemes as evidenced by their use in gravitational wave detectors \cite{grote2007high}, cavity-stabilised lasers \cite{Robinson2019}, and the generation of entangled states \cite{serikawa2018_prl}. 

\begin{figure}[ht]
    \centering
    \includegraphics[width=0.45\textwidth]{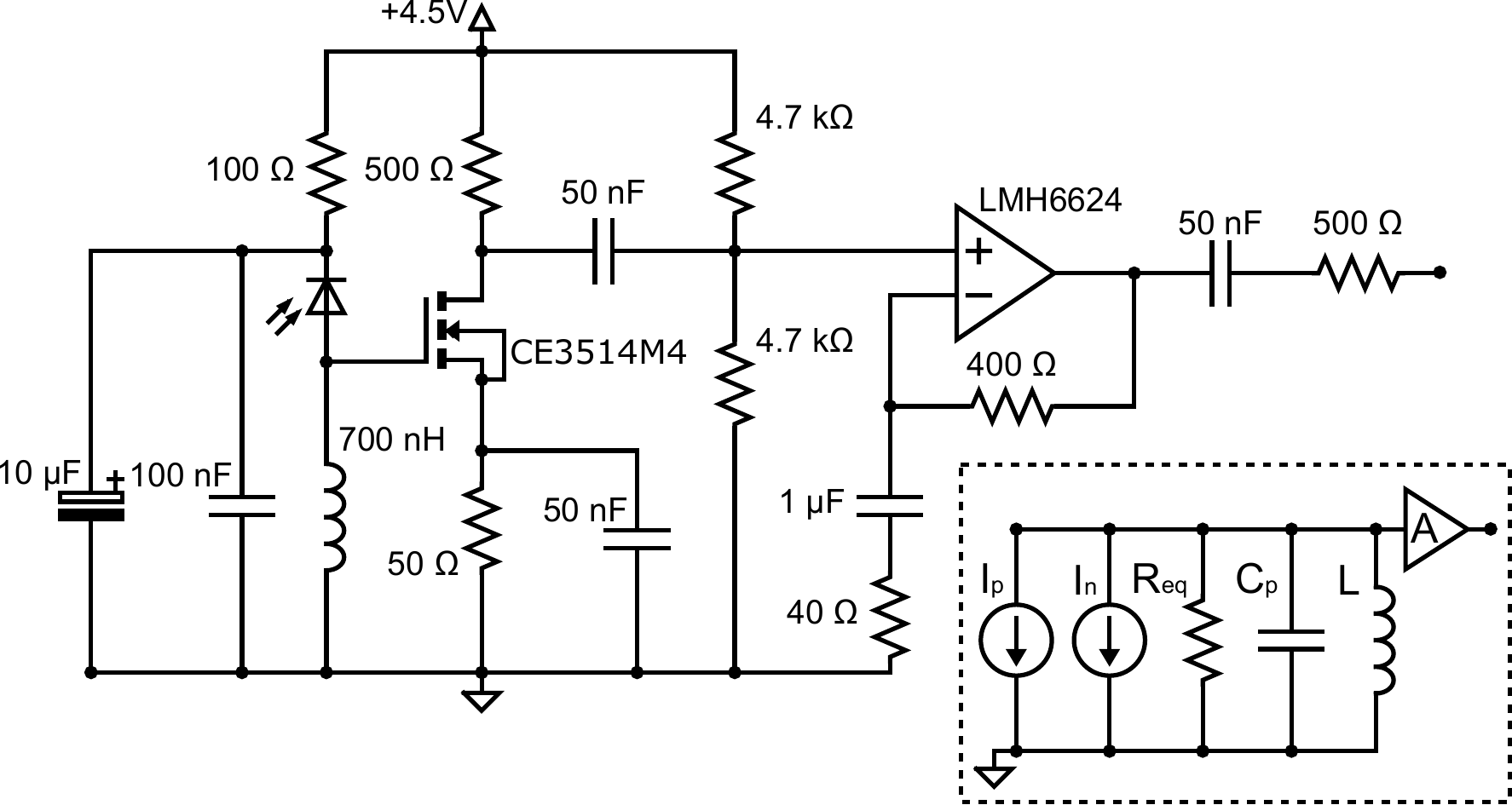}
    \caption{ \label{fig:circuit} Circuit diagram showing the two-stage resonant TIA photodetector. Not shown is the on-board voltage reference used to generate the supply voltage and the op-amp decoupling capacitors. Inset: the simplified circuit used for modeling which includes the photocurrent $I_p$ and the input current noise $I_n$ in parallel with the the effective transconductance $R_{eq}$ set by the gate resistance (see equation \ref{eq:R}), the inductance $L$, and the input capacitance $C_p$---arising primarily from the photodiode.}
\end{figure}

The resonant photodetector circuit, which builds on several previous designs \cite{chen2016, eberle2013, steinlechner2013}, is shown in figure \ref{fig:circuit}. The main modification presented in this work is the inclusion of a JFET common-source pre-amplifier at the input located between the diode and op-amp. Without this buffer amplifier, the photodetector exhibited high frequency noise which we attributed to excess op-amp input current noise. Initial prototype designs that had the resonant input connected directly to the non-inverting op-amp stage were unable to achieve an input-referred current noise below 3 $\textrm{pA}/\sqrt{\textrm{Hz}}$ at \SI{50}{\mega\hertz} or higher\cite{Note}. One option would be to use a transformer \cite{chen2017} to couple the diode to the amplifier in order to boost the photocurrent, but this has the drawback of also amplifying the diode capacitance \cite{potnis2016note}. With the JFET input, the current noise depends on the drain leakage current set by the effective resistance $R_{eq}$ of the gate. This is given by the real part of the frequency dependent input impedance of the JFET gate \cite{motchenbacher1993}, equal to:

\begin{equation}
 \label{eq:R}
    R_{eq} = \frac{1+\omega^2R_L^2C_L^2}{\omega^2g_mR_L^2C_LC_{gd}},
\end{equation}

\noindent where $R_L$ and $C_L$ are the output load resistance and capacitance, $C_{gd}$ is the parasitic capacitance coupling the drain to the gate, and $g_m$ is the JFET transconductance. This resistance gives rise to a thermal Johnson noise current $I_t = \sqrt{4k_bT/R}$ which sets the minimum achievable NEP. Therefore, it is advantageous to choose a JFET with small parasitic capacitance in order to achieve a large transimpedance. One may also want to choose a JFET with small transconductance and low output resistance, but it is important to recognize that this reduces the small signal gain of the amplifier, thus increasing the input referred voltage noise of the op-amp. 

The complex transimpedance $Z(s)$ of the photodetector is given by:

\begin{align}
Z(s) = \frac{sL}{(1+s^2LC_p)+sL/R_{eq}}&\\
=R_{eq}\frac{s/\omega_0Q}{1+s/\omega_0Q+s^2/\omega_0^2},
\end{align}

\noindent which is set by the parallel combination of the gate resistance, the input capacitance $C_p$ and the inductance \cite{InductorNote} $L$. As the values $C_p$ and $R_{eq}$ are not precisely known, it is useful to express the transimpedance in terms of the resonant frequency ($\omega_0 = 1/\sqrt{LC_p}$) and the Q-factor ($Q = R_{eq}\sqrt{C/L}$) which can both be directly measured.

The JFET used for the circuit was the CE3521M4 \cite{someotherNote}, manufactured by California Eastern Laboratories, which is similar to the JFET used in other resonant photodetectors \cite{serikawa2018}. It was chosen based on its low noise-factor and high transconductance. Unfortunately, no information is provided regarding its input and reverse transfer capacitance. As a result, it is not possible to verify equation \ref{eq:R}. However, its intended application for high-frequency microwave amplifiers requires such capacitance values to be small. The op-amp used was the LMH6624, manufactured by Texas Instruments. Its high gain-bandwidth product, low input noise, and small capacitance makes it well-suited for this application. It was observed that the NEP could be improved by operating the op-amp slightly below its recommended minimum supply level of \SI{5}{\volt}.  The PIN photodiode used for the detector was the Hamamatsu S5973-02 which has a responsivity $A_R$ equal to \SI{0.34}{\ampere\per\watt} and a measured capacitance of 2 pF. Care is taken to minimize further PCB related parasitic capacitance by removing the ground planes under the signal path to ensure that the input capacitance is dominated by the diode’s contribution. The entire circuit is powered through an on-board voltage regulator and is enclosed in a metal case.

The detector's frequency response and NEP was characterised by measuring the shot noise limited amplitude noise present on an input signal of known optical power $P_i$. The resulting shot noise on the photocurrent $I_s$ is given by $\sqrt{2qA_RP_i}$, where $q$ is the fundamental unit of charge. To boost the signal strength, an additional amplifier (Mini-circuits ZFL-500LN+) was added at the output. In general, if the incident light increases the power spectral density noise by $\Delta$ dB, the incident power $P_{\textrm{NEP}}$ which will induce a current shot noise equivalent to the detector current noise is given by:

\begin{equation}
P_{\textrm{NEP}} = \frac{P_i}{10^{\Delta/10}-1}
\end{equation}

\begin{figure}[ht]
    \centering
    \includegraphics[width=0.5\textwidth]{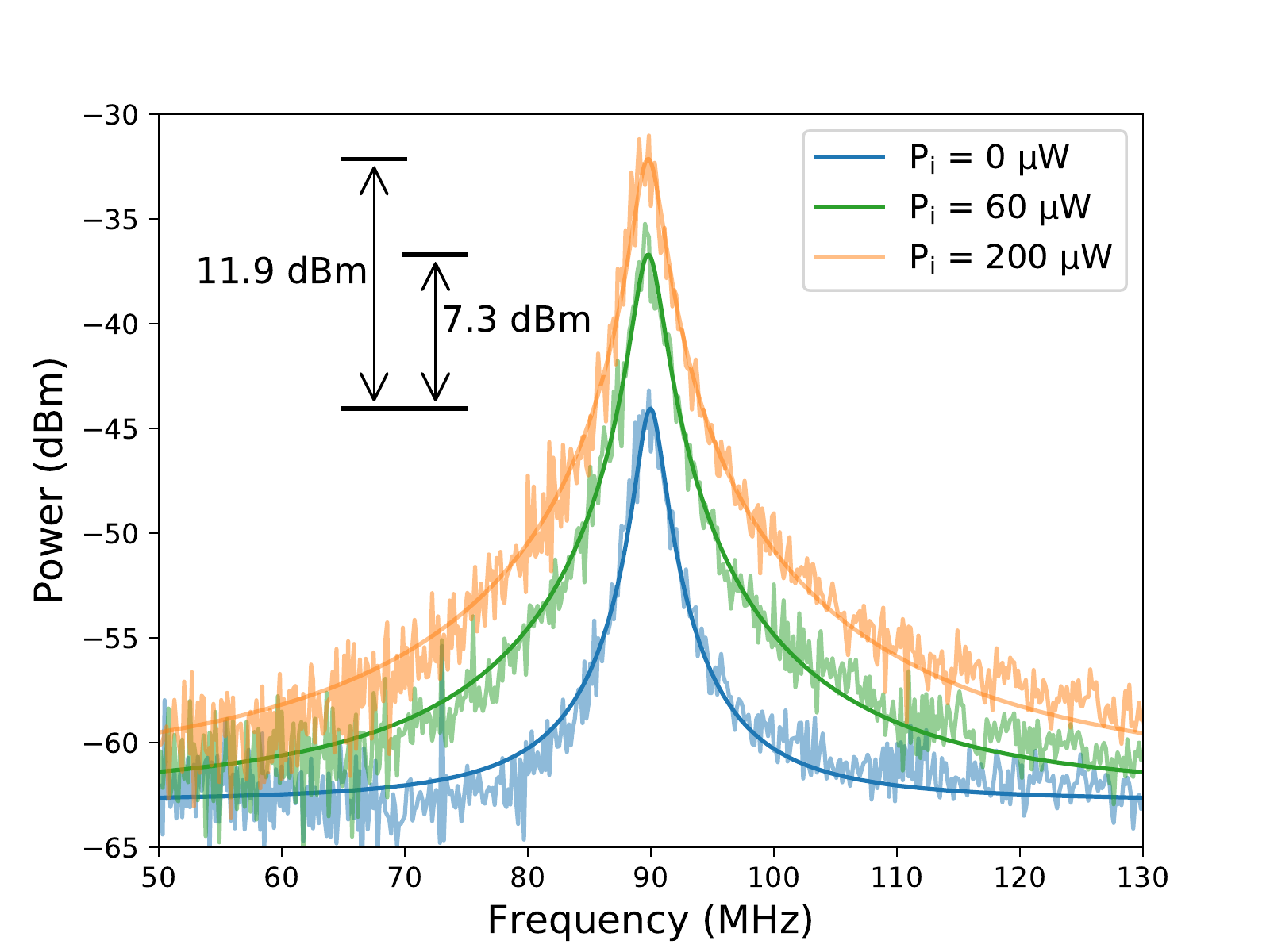}
    \caption{ \label{fig:NEP} Output spectrum of the resonant photodetector with and without input light. The increase in noise can be attributed to the inherent amplitude shot-noise present on the light which can be used to determine the NEP of the detector. Not shown is the detection noise floor, which is well below the measured photodetector noise. The above spectra are an average of 10 sweeps with a resolution bandwidth of \SI{30}{\kilo\hertz}.}
\end{figure}

The light source used for the tests is an intensity stabilized external cavity diode laser. The laser was verified to have less than 1~dB of excess intensity noise between 10 and 100~MHz by comparing the response of a photodiode in balanced vs unbalanced configuration. However, in order to suppress the excess intensity noise further, the light is filtered by a Fabry-P\'erot cavity with a full-width at half-maximum of 20~MHz. At incident powers of 60 and \SI{200}{\micro\watt}, the peak output noise increased by 7.3 and 11.9 dB respectively, as shown in figure \ref{fig:NEP}. Both measurements yield a $P_\textrm{NEP}$ of \SI{14}{\micro\watt} corresponding to equivalent input referred current noise of 1.2 $\textrm{pA}/\sqrt{\textrm{Hz}}$. From the width of the noise spectra, the Q-factor of the resonance was measured to be 39, yielding an effective transimpedance of \SI{16}{\kilo\ohm}. The resulting Johnson current noise associated with this resistance is 1.0~ $\textrm{pA}/\sqrt{\textrm{Hz}}$, indicating that there is an excess current noise of 0.7~$\textrm{pA}/\sqrt{\textrm{Hz}}$. The transfer function of the photodetector was independently verified by using an amplitude modulated beam and measuring the circuit's response. By changing the inductor value, the measurement was repeated for two other resonant frequencies, \SI{77}{\mega\hertz} ($L$ = \SI{1}{\micro\henry}) and \SI{123}{\mega\hertz} ($L$ = \SI{0.5}{\micro\henry}). The resulting input current noise values were measured to be 1.0~$\textrm{pA}/\sqrt{\textrm{Hz}}$ and 1.4~$\textrm{pA}/\sqrt{\textrm{Hz}}$ respectively, consistent with the fact that the effective transimpedance is reduced at high frequencies.  

In an attempt to verify equation \ref{eq:R}, the JFET was switched to an IF142, manufactured by InterFET, as the required parameters needed to calculate the gate resistance are given in the datasheet ($g_m$~$=$~$3$~mS, $C_{gd}$~$=$~$0.6$~pF). Combined with the load resistance of \SI{400}{\ohm} and capacitance of 2~pF set by the op-amp input terminals, the estimated gate resistance at the resonant frequency is \SI{5}{\kilo\ohm}. Based on the measured Q-value of the detector's response, the gate resistance is approximately \SI{4.5}{\kilo\ohm}. The resulting NEP was measured to be 5.3 $\textrm{pW}/\sqrt{\textrm{Hz}}$ which corresponds to an input referred current noise of 1.8 $\textrm{pA}/\sqrt{\textrm{Hz}}$---in reasonable agreement with the value of 1.5 $\textrm{pA}/\sqrt{\textrm{Hz}}$ set by the Johnson noise of the predicted gate resistance.

In conclusion, we have presented a simple design for a low noise resonant photodetector and characterised its performance. We have presented a simple model which can be used to predict the effective transimpedance of the photodetector set by the JFET's gate resistance. The input current noise of 1.2 $\textrm{pA}/\sqrt{\textrm{Hz}}$ at \SI{90}{\mega\hertz} is near the theoretical limit imposed by the Johnson noise set by the transimpedance and corresponds to shot noise limited performance for incident optical powers above \SI{14}{\micro\watt}. The intended use of this photodetector is for cavity-based non-destructive detection of trapped strontium atoms \cite{Vallet2017}, but it can be easily modified to operate at a different resonant frequency or wavelength, making it well suited to a range of applications.

This work was funded by the UK Department for Business, Energy and Industrial Strategy as part of the National Measurement System Programme. The authors thank Jochen Kronjaeger for useful discussions and designing the photodiode enclosure. Figure 1 was made using Circuit Diagram, an online open source program for drawing circuits. 

\nocite{*}
\bibliography{aipsamp}

\begin{thebibliography}{15}
\bibitem [{Not()}]{Note}%
  \BibitemOpen
  \href@noop {} {}\bibinfo {note} {Op-amp input current noise, especially for
  FET inputs, climbs quickly with frequency and can exceed the nominal value
  given in the datasheet. Unfortunately, high frequency current noise is rarely
  specified in most data-sheets except for rare examples, e.g. the LTC6268
  which has a current noise of 7 $\textrm{fA}/\sqrt{\textrm{Hz}}$ at
  \SI{100}{\kilo\hertz}, but over 5 $\textrm{pA}/\sqrt{\textrm{Hz}}$ at
  \SI{100}{\mega\hertz}.}\BibitemShut {Stop}%
\bibitem [{Ind()}]{InductorNote}%
  \BibitemOpen
  \href@noop {} {}\bibinfo {note} {We used an air coil inductor manufactured by
  Coilcraft (part number 2929SQ-501GEC). It was chosen based on its high self
  resonant frequency (\SI{485}{\mega\Hz}). Furthermore, the inductance can be
  changed by splaying out the coil windings, allowing for fine tuning of the
  resonance frequency by several megahertz.}\BibitemShut {Stop}%
\bibitem [{som()}]{someotherNote}%
  \BibitemOpen
  \href@noop {} {}\bibinfo {note} {Commercial components are identified for the
  sake of clarity and does not imply that they are necessarily the best
  alternatives.}\BibitemShut {Stop}%
\end{thebibliography}

\end{document}